\documentstyle[12pt]{article}

\begin{document}
\vspace*{2cm}
\centerline {\huge Nonlinear Image Processing Based}
\medskip
\centerline {\huge on Optimization of Generalized}
\medskip
\centerline {\huge Information Methods}

\vspace*{12mm}

\centerline{Anisa T. BAJKOVA}

\centerline{\it IAA RAS, Nab. Kutuzova, 10, 191187, St-Petersburg, Russia}

\vspace*{7mm}

{\baselineskip 4pt
\begin{quote}
\begin{quote}
\noindent
{\small
{\bf Abstract}.
A range of nonlinear image reconstruction procedures based on extremizing
the generalized Shannon entropy, Kullback-Leibler cross-entropy and Renyi
information measures and proposed by the author in early papers is presented.
The ``generalization'' assumes search for the solution over the space
of real bipolar or complex functions. Such an approach allows, first, to
reconstruct signals of any type and physical nature and, secondly,
to decrease nonlinear intensity image distortions caused by measurement
errors. All the elaborated procedures are contained in VLBI ``IMAGE'' program
package developed in IAA RAS.
}
\end{quote}
\end{quote}
}
\vspace*{7mm}

\noindent {\bf 1.Introduction}

\vspace*{6mm}

The main aim of this paper is consideration of a class of image
deconvolution (reconstruction) algorithms based on optimization
of different information measures. The most popular among them
is the well known maximum entropy method in Shannon formulation.
The generalized information methods based on extremizing entropic
(Shannon, Kullback-Leibler) and Renyi measures were proposed
by the author in several previous publications [1--3] for
reconstruction of real bipolar and complex images. In this paper we give
a brief review of the proposed techniques.

All the proposed methods are realized in "IMAGE" program
package elaborated in IAA RAS in framework of "QUASAR" Russian Very
Long Baseline Interferometry (VLBI) project.

\vspace*{9mm}

\noindent{\bf 2.Entropic Processing}

\vspace*{6mm}

\noindent{\bf 2.1.Maximum Entropy Method}

\vspace*{6mm}

Maximum entropy method (MEM) is widely used for image reconstruction from
incomplete and noisy Fourier data. In Fourier space the image reconstruction
problem is equivalent to one of extrapolation and interpolation.
The MEM produces maximally smooth, or unbiased, estimates [4]. If $r_{ml}$ is
a real non-negative signal (we consider two-dimensional sampling signals)
to be estimated then the maximum entropy method assumes solving the
following optimization problem with linear constraints dictated by data

\begin{equation}
\max\sum_{m}\sum_{l}{-(r_{ml}\ln(r_{ml}))}-1/2\sum_{n}\sum_{k}(\eta^{r2}_{nk}+\eta^{i2}_{nk})/\sigma^2_{nk},~~r_{ml}\ge0,
\end{equation}

\begin{equation}
\sum_m\sum_l{r_{ml}a_{ml}^{nk}}+\eta^r_{nk}=A_{nk},~~~\sum_m\sum_l{r_{ml}b_{ml}^{nk}}+\eta^i_{nk}=B_{nk},
\end{equation}
where $a_{ml}^{nk}, b_{ml}^{nk}$ are constants determined by linear image
formation system (Fourier transform), $A_{nk}, B_{nk}$ are the real and
imaginary parts of the Fourier data respectively, $\eta^r_{nk}, \eta^i_{nk}$
are noise components of data, $\sigma^2_{nk}$ are noise variances for
each of the real and imaginary parts at sample $nk$. Note that the signal
must be real and non-negative-definite for a real entropy functional to
exist.

Using the Lagrange method it is easy to obtain a solution for $r_{ml}$
and noise components
\begin{equation}
r_{ml}=\exp(-\sum_n\sum_k{\alpha_{nk}a_{ml}^{nk}+\beta_{nk}b_{ml}^{nk}-1}),
\end{equation}

\begin{equation}
\eta^r_{nk}=\sigma^2_{nk}\alpha_{nk},~~\eta^i_{nk}=\sigma^2_{nk}\beta_{nk},
\end{equation}
where $\alpha_{nk}, \beta_{nk}$ are conventional Lagrange multipliers.

\vspace*{6mm}
\noindent{\bf 2.2.Kullback-Leibler Entropy Method}
\vspace*{6mm}

Kullback-Leibler information permits insertion of a prior bias function
$a_{ml}$ into entropy functional (1) as follows
\begin{equation}
\sum_m\sum_l{r_{ml}\ln(r_{ml}/a_{ml})},~~r_{ml},~a_{ml}\ge 0.
\end{equation}
The aim of such insertion is improving the entropic algorithm (1) using
a priori information about an unknown signal [2]. Usually as a bias function
$a_{ml}$ a first approximation to $r_{ml}$ obtained by inverse Fourier
transform of input Fourier data is used. Missing data regions in Fourier
space are filled out by zeros.

\vspace*{9mm}

\noindent{\bf 3.Renyi Information Processing}

\vspace*{6mm}

Another powerful method of reconstruction considered here is based on
minimization of the following Renyi information measure [3]
\begin{equation}
\sum_m\sum_l{r_{ml}^{\alpha}a_{ml}^{1-\alpha}},~~r_{ml},a_{ml}\ge 0,~~ \alpha\ne 0,
\end{equation}
where $a_{ml}$ is a reference function.

\vspace*{9mm}

\noindent{\bf 4.Generalized Methods}

\vspace*{6mm}

Originally the generalized maximum entropy method (GMEM) was proposed by
Bajkova [1] for reconstruction of complex coherent images formed by radio
holography principle. In [2] the GMEM idea was used by Frieden for
generalization of Kullback-Leibler cross-entropy reconstruction of ISAR
images. In [3] by Frieden the generalized form of entropy functionals
was extended to minimum Renyi information method. Bajkova proposed the
method of solution of corresponding generalized optimization problem [3].

\vspace*{6mm}

\noindent{\bf 4.1.Generalized Maximum Entropy Method}

\vspace*{6mm}

Here let us recall the basic aspects of the GMEM [1].

First, consider a real signal $r_{ml}$ which may take both positive and
negative values. In this case the minimizing functional cannot be
written as (1). Therefore it was proposed in [1] to represent entropy (1)
as entropy of the absolute value of $r_{ml}$:

\begin{equation}
\sum_m\sum_l{|r_{ml}|\ln(|r_{ml}|)}.
\end{equation}

To represent the optimization problem in traditional form (1)-(2) let us
represent the signal sought for  $r_{ml}$ as a difference between two
non-negative-definite functions:

\begin{equation}
r_{ml}=x_{ml}-y_{ml},~~ x_{ml}\ge 0,~~ y_{ml}\ge 0.
\end{equation}

Thus the positive regions are represented by $x_{ml}$ and the
negative by $y_{ml}$. Curves $x_{ml}$ and $y_{ml}$ do not have overlapping
support regions.
Nonoverlapping condition can be written as
\vskip 6mm
\centerline{if $r_{ml}>0$ then $y_{ml}\rightarrow 0$ and $r_{ml}=x_{ml}$,}
\vskip 3mm
\centerline{if $r_{ml}<0$ then $x_{ml}\rightarrow 0$ and $r_{ml}=-y_{ml}$.}
\vskip 6mm
Then the problem of estimation of $r_{ml}$ can be
replaced by one of estimation of $x_{ml}$ and $y_{ml}$ as
\begin{equation}
\max\sum_m\sum_l{-(x_{ml}\ln(x_{ml})+y_{ml}\ln(y_{ml}))}-1/2\sum_n\sum_k(\eta^{r2}_{nk}+\eta^{i2}_{nk})/\sigma^2_{nk}.
\end{equation}

Jumping to a complex signal $u_{ml}=r_{ml}+jq_{ml},~j=\sqrt -1$, where
both $r_{ml}$ and $q_{ml}$ are real bipolar signals, we can represent
similarly to (8)
the sought for sequences as the differences of non-negative-definite
sequences as well:
\begin{equation}
r_{ml} = x_{ml}-y_{ml},~~~ q_{ml} = z_{ml}-v_{ml},
\end{equation}
\noindent
where $x_{ml}\ge 0,~ y_{ml}\ge 0,~ z_{ml}\ge 0,~ v_{ml}\ge 0$.

Obviously, if the sequences $x_{ml},y_{ml}$ and
$z_{ml},v_{ml}$ do not overlapp, then $x_{ml}$ and $z_{ml}$ determine
positive parts and $y_{ml}$ and $v_{ml}$ determine negative parts of the
sequences $r_{ml}$ and $q_{ml}$ respectively.

Nonoverlapping condition mentioned above must be complemented by one for
the signal $q_{ml}$:
\vskip 6mm
\centerline{if $q_{ml}>0$ then $v_{ml}\rightarrow 0$ and $q_{ml}=z_{ml}$,}
\vskip 3mm
\centerline{if $q_{ml}<0$ then $z_{ml}\rightarrow 0$ and $q_{ml}=-v_{ml}$.}
\vskip 6mm

Then the minimized entropic functional for estimation of the complex signal
can be written as the following functional
\begin{equation}
\sum_m\sum_l{x_{ml}\ln(\alpha x_{ml})+y_{ml}\ln(\alpha y_{ml})}+z_{ml}\ln(\alpha z_{ml})+v_{ml}\ln(\alpha v_{ml})
\end{equation}
with inserted a positive parameter $\alpha$ responsible for not overlapping
positive and negative parts of the search for real bipolar sequences $r_{ml}$
and $q_{ml}$. Significance of the parameter $\alpha$ will be clear below from
equations (15)-(17).

Linear constraints (2) derived from Fourier data must be rewritten in the
way:
\begin{equation}
\sum_m\sum_l{(x_{ml}-y_{ml})a^{nk}_{ml}-(z_{ml}-v_{ml})b^{nk}_{ml}}+\eta^r_{nk} = A_{nk},
\end{equation}
\begin{equation}
\sum_m\sum_l{(x_{ml}-y_{ml})b^{nk}_{ml}+(z_{ml}-v_{ml})a^{nk}_{ml}}+\eta^i_{nk}
= B_{nk},
\end{equation}
\begin{equation}
{x_{ml}\ge 0,~~ y_{ml}\ge 0,~~ z_{ml}\ge 0,~~ v_{ml}\ge 0},
\end{equation}
\noindent
where $a^{nk}_{ml}, b^{nk}_{ml}$ are constant coefficients which are determined
by Fourier transform, $A_{nk}, B_{nk}$ are measured real
and imaginary parts of Fourier data respectively.
The solutions of optimization problem (11)--(14) obtained using the Lagrange
method for signals
$(x_{ml},y_{ml},z_{ml},v_{ml})$ look like
\begin{equation}
x_{ml}=\exp(-\sum_n\sum_k{\alpha_{nk}a_{ml}^{nk}+\beta_{nk}b_{ml}^{nk}-1-\ln\alpha}),
\end{equation}
\begin{equation}
v_{ml}=\exp(-\sum_n\sum_k{\alpha_{nk}b_{ml}^{nk}-\beta_{nk}a_{ml}^{nk}-1-\ln\alpha}),
\end{equation}
\begin{equation}
x_{ml}y_{ml} = z_{ml}v_{ml} = \exp(-2-2\ln\alpha).
\end{equation}
\noindent
Making parameter $\alpha$ larger we can reach not overlapping effect.
Usual value of $\alpha$ is 1000.

Unknown Lagrange multipliers can be found from the following dual optimization
problem without supplementary conditions

\begin{eqnarray}
&&{\max\sum_m\sum_l{-(x_{ml}+y_{ml}+z_{ml}+v_{ml})}-1/2\sum_n\sum_k(\eta^{r2}_{nk}+\eta^{i2}_{nk})/\sigma^2_{nk}}
\nonumber \\
&&{-\sum_n\sum_k{(\alpha_{nk}A_{nk}+\beta_{nk}B_{nk})}-\sum_n\sum_k{(\alpha_{nk}\eta^r_{nk}+\beta_{nk}\eta^i_{nk}})}
\end{eqnarray}
by substituting expressions (15)-(17) for $x_{ml}, y_{ml}, z_{ml}, v_{ml}$
and (4) for $\eta^r_{nk}, \eta^i_{nk}$.

Equations (15)-(17) show that for any real solution $\{\alpha_{nk}\},\{\beta_{nk}\}$
to the problem, necessarily
\begin{equation}
x_{ml}\ge 0,~~y_{ml}\ge 0,~~z_{ml}\ge 0,~~v_{ml}\ge 0.
\end{equation}
That is all the reconstructed signals are non-negative. This is
important because non-negative-constrained solutions, being nonlinear in the
data, can have higher bandwidth and hence higher resolution than the data.
By (10) this is now true for real signals $r_{ml}$ and $q_{ml}$ and hence
for {\it complex} signal $u_{ml}$ as well.

So, the GMEM approach allows to obtain a solution of the reconstruction
problem in the space of complex functions. Because of non-negativity of the
solutions (15)-(17) the method is nonlinear and possesses a super resolution
effect similarly to the classical maximum entropy method.

\vspace*{6mm}

\noindent{\bf 4.2.Generalized Kullback-Leibler Method}

\vspace*{6mm}

Generalized Kullback-Leibler method assumes prior knowledge of bias
non-negative functions $a_{ml}, b_{ml}, c_{ml}, d_{ml}$ corresponding to
the signals $x_{ml}, y_{ml}, z_{ml}, v_{ml}$ respectively. Then the
entropic functional (11) is modified as
\begin{eqnarray}
&&{\sum_m\sum_l{x_{ml}\ln(\alpha x_{ml}/a_{ml})+y_{ml}\ln(\alpha y_{ml}/b_{ml})}}
\nonumber \\
&&{+z_{ml}\ln(\alpha z_{ml}/c_{ml})+v_{ml}\ln(\alpha v_{ml}/d_{ml})}.
\end{eqnarray}

In this case, bias functions can be again the smoothed inverse Fourier
transforms of Fourier data with zeros in the points where measurements
are absent.

Using data constraints (12)-(13) and the Lagrange method we obtain the
following solution to sought for non-negatively determined signals
\begin{equation}
x_{ml}=(a_{ml}/e\alpha)\exp(-\sum_n\sum_k{\alpha_{nk}a_{ml}^{nk}+\beta_{nk}b_{ml}^{nk}}-1),
\end{equation}
\begin{equation}
v_{ml}=(d_{ml}/e\alpha)\exp(-\sum_n\sum_k{\alpha_{nk}b_{ml}^{nk}-\beta_{nk}a_{ml}^{nk}}-1),
\end{equation}
\begin{equation}
x_{ml}y_{ml} = a_{ml}b_{ml}e^{-2}\alpha^{-2},
\end{equation}
\begin{equation}
z_{ml}v_{ml} = c_{ml}d_{ml}e^{-2}\alpha^{-2}.
\end{equation}

As seen, maximizing Kullback-Leibler entropy functional subject to
the linear data constraints (12)-(13) produces solutions (21)-(24) that are
linearly proportional to the input functions $a_{ml},b_{ml},c_{ml},d_{ml}$
respectively.

\vspace*{6mm}

\noindent{\bf 4.3.Generalized Minimum Renyi Information Method}

\vspace*{6mm}

Similarly to maximum entropy method for complex
signal $u_{ml}=r_{ml}+jq_{ml}$ we suggest [3] to minimize the sum of two
Renyi measures of absolute values of $r_{ml}$ and $q_{ml}$. Then
functional (6)is modified as
\begin{equation}
\sum_m\sum_l{-(|r_{ml}|^{\alpha}a_{ml}^{1-\alpha}+|q_{ml}|^{\alpha}b_{ml}^{1-\alpha})},
\end{equation}
where $a_{ml}$ and $b_{ml}$ are input modulus reference signals.

Representing the sought for sequences in the form of differences between
positive and negative parts as in (10) we can rewrite previous
functional (25)
as follows
\begin{equation}
\sum_m\sum_l{-(x_{ml}^{\alpha}a_{ml}^{1-\alpha}+y_{ml}^{\alpha}b_{ml}^{1-\alpha}}+z_{ml}^{\alpha}c_{ml}^{1-\alpha}+v_{ml}^{\alpha}d_{ml}^{1-\alpha}),
\end{equation}
where $a_{ml},b_{ml},c_{ml},d_{ml}$ are corresponding reference signals.

Direct using the Lagrange method is very complicated. In order to simplify
solution let us make the following substitution
$\alpha=1+1/2\mu$ and introduce the following new variables:
\begin{equation}
t_{ml}(s_{ml},h_{ml},g_{ml})=(x_{ml}(s_{ml},h_{ml},g_{ml})/a_{ml}(b_{ml},c_{ml},d_{ml}))^{1/2\mu}.
\end{equation}

Then the optimization problem (20), (12)-(14) can be rewritten with respect
to new variables (27) as
\begin{eqnarray}
&&{\max\sum_m\sum_l{-(a_{ml}t_{ml}^{2\mu+1}+b_{ml}s_{ml}^{2\mu+1}}+{c_{ml}h_{ml}^{2\mu+1}+d_{ml}g_{ml}^{2\mu+1})}}
\nonumber \\
&&{-1/2\sum_{n}\sum_{k}(\eta^{r2}_{nk}+\eta^{i2}_{nk})/\sigma^2_{nk}},
\end{eqnarray}
\begin{equation}
\sum_m\sum_l{(a_{ml}t_{ml}^{2\mu}-b_{ml}s_{ml}^{2\mu})a_{ml}^{nk}}-{(c_{ml}h_{ml}^{2\mu}-d_{ml}g_{ml}^{2\mu})b_{ml}^{nk}}+\eta^r_{nk}=A_{nk},
\end{equation}
\begin{equation}
\sum_m\sum_l{(a_{ml}t_{ml}^{2\mu}-b_{ml}s_{ml}^{2\mu})b_{ml}^{nk}}+{(c_{ml}h_{ml}^{2\mu}-d_{ml}g_{ml}^{2\mu})a_{ml}^{nk}}+\eta^i_{nk}=B_{nk},
\end{equation}
\begin{equation}
t_{ml}\ge0,~~ s_{ml}\ge0,~~ h_{ml}\ge0,~~ g_{ml}\ge0.
\end{equation}

Using the Lagrange method we find the following solution for sought for
signals:
\begin{equation}
t_{ml}=-2\mu/(2\mu+1)\sum_n\sum_k{\alpha_{nk}a_{ml}^{nk}+\beta_{nk}b_{ml}^{nk}},
\end{equation}
\begin{equation}
h_{ml}=2\mu/(2\mu+1)\sum_n\sum_k{\alpha_{nk}b_{ml}^{nk}-\beta_{nk}a_{ml}^{nk}},
\end{equation}
\begin{equation}
t_{ml}=-s_{ml};~~ h_{ml}=-g_{ml}.
\end{equation}
where $\alpha_{nk}$ and $\beta_{nk}$ are Lagrange multipliers.

Remembering condition (31) and taking (34) we can say that the negative and positive
parts of sought for real solutions to $r_{ml}$ and $q_{ml}$ never overlapp
and no parameter (as $\alpha$ in the GMEM) is required.
That is an advantage of the generalized minimum Renyi information method.

Sought for a complex signal $u_{ml}$ can be found by inverting equations
(27) and using representation (10).

It is necessary to note that the quality of image reconstruction by Renyi
method depends on choosing parameter $\alpha$ in the functional.
When $\alpha=2 (\mu=0.5)$ we have a conventional minimum intensity
criterion. Experimentally established that when $\alpha$ approaches $1$
by increasing $\mu$ we obtain a criterion with more strong nonlinear
(extrapolation) features [5].

\vspace*{9mm}

\noindent{\bf 5.A problem of nonlinear distortions}

\vspace*{6mm}

Let us consider real non-negative images.
Because of nonlinearity nature of considered above deconvolution procedures
there is a problem connected with nonlinear image distortions due to noise in
spectral data [6]. We investigated only the case of additive noise in data.
As experiments show the generalized algorithms ensure much less artefacts
than the classical ones. This phenomenon can be explained that
noise can so degrade data that to the latter in general a real solution
with both positive and negative values will correspond
but not pure real non-negative one which is sought by conventional
algorithms. Therefore we consider that seeking for
a solution of image reconstruction problem in generalized form is more adequate.

\vspace*{9mm}

\noindent{\bf 6.The "IMAGE" program package}

\vspace*{6mm}

"IMAGE" program package elaborated in IAA RAS [7] in framework of
"QUASAR" VLBI project is intended for VLBI mapping natural (incoherent)
and artificial (coherent, ISAR) radio sources and investigation
of new image reconstruction procedures.
At present the "IMAGE" program package contains except of the CLEAN and
classical information nonlinear methods a number of generalized versions of
information methods.
For solution of related optimization problems both steepest descent
and Newton-Raphson numerical methods are realized.

\vspace*{9mm}

\noindent{\bf 7.Conclusion}

\vspace*{6mm}

Maximum entropy, maximum Kullback-Leibler cross-entropy
and minimum Renyi information methods having good extrapolation and
interpolation features are
generalized for reconstruction of signals of any type including {\it
complex} ones. They may be used
for reconstruction of real bipolar and complex signals or reconstruction
of real
non-negative images with minimal nonlinear distortions.

This work was supported by Russian Foundation for Basic Researches under
grant N 96-02-19177.

\vspace*{9mm}

{\baselineskip 4pt
\noindent{\small {\bf References}

\vspace*{4.5mm}
\hangindent=1cm \noindent
[1] \hskip 4.5mm Bajkova A. The generalization of maximum entropy method for
reconstruction of complex functions.
{\it Astron. \& Astroph. Tr.,} {\bf 1} (1992), 313--320.

\hangindent=1cm \noindent
[2] \hskip 4.5mm Frieden R., Bajkova A. Bayesian cross-entropy reconstruction of complex
images.
{\it Applied Optics,} {\bf 33} (Jan. 1994), 219--226.

\hangindent=1cm \noindent
[3] \hskip 4.5mm Frieden R., Bajkova A. Reconstruction of complex signals using minimum
Renyi information.
{\it Applied Optics,} {\bf 34} (July 1995), 4086--4093.

\hangindent=1cm \noindent
[4] \hskip 4.5mm Frieden R. Restoring with maximum likelihood and maximum entropy.
{\it J. Opt. Soc. Am.,} {\bf 62} (1972), 511--518.

\hangindent=1cm \noindent
[5] \hskip 4.5mm Bajkova A. Renyi measure for image reconstruction in Astronomy.
{\it Communications of IAA RAS,} No 59, 1994.

\hangindent=1cm \noindent
[6] \hskip 4.5mm Bajkova A. On the problem of nonlinear distortions of the maximum
entropy method.
{\it Izvestia Vuzov. Radiofizika,} {\bf 38} (1994), 1267--1277.

\hangindent=1cm \noindent
[7] \hskip 4.5mm Bajkova A. New version of ``Image'' program package in
Linux for modeling and processing VLBI images.
       {\it Communications of IAA RAS}, No 120, 1998.

}
}
\end{document}